# Strong bulk-surface interaction dominated in-plane anisotropy of electronic structure in GaTe

Kang Lai[1], Sailong Ju[1], Hongen Zhu[2], Hanwen Wang[3], Hongjian Wu[1], Bingjie Yang[4], Enrui Zhang[1], Ming Yang[1], Fangsen Li[4], Shengtao Cui[2], Xiaohui Deng[5], Zheng Han[6,7]*, Mengjian Zhu[8]*, Jiayu Dai[1]*

[1]Department of Physics, National University of Defense Technology, Changsha 410073, China

[2]National Synchrotron Radiation Laboratory, University of Science and Technology of China, Hefei 230029, China

[3]Shenyang National Laboratory for Materials Science, Institute of Metal Research, Chinese Academy of Sciences, Shenyang 110016, China

[4]School of Nano-Tech and Nano-Bionics, University of Science and Technology of China, Hefei 230026, China

[5]College of Physics and Electronic Engineering, Hengyang Normal University, Hengyang 421002, China

[6]State Key Laboratory of Quantum Optics and Quantum Optics Devices, Institute of Opto-Electronics, Shanxi University, Taiyuan 030006, China

[7]Collaborative Innovation Center of Extreme Optics, Shanxi University, Taiyuan 030006, China

[8]College of Advanced Interdisciplinary Studies, National University of Defense Technology, Changsha 410073, China

*Corresponding to Z.H. (email: vitto.han@gmail.com) or to M.Z. (email: zhumengjian11@nudt.edu.cn) or to J.D. (email: jydai@nudt.edu.cn).

**Recently, intriguing physical properties have been unraveled in anisotropic layered semiconductors, in which the in-plane electronic band structure anisotropy often originates from the low crystallographic symmetry and thus a thickness-independent character emerges. Here, we apply high-resolution angle-resolved photoemission spectroscopy to directly image the in-plane anisotropic energy bands in monoclinic gallium telluride (GaTe). Our first-principles calculations reveal the in-plane anisotropic energy band structure of GaTe measured experimentally is dominated by a strong bulk-surface interaction rather than geometric factors, surface effect and quantum confinement effect. Furthermore, accompanied by the thickness of GaTe increasing from mono- to few-layers, the strong interlayer coupling of GaTe induces direct-indirect-direct band gap transitions and the in-plane anisotropy of hole effective mass is reversed. Our results shed light on the physical origins of in-plane anisotropy of electronic structure in GaTe, paving the way for the design and device applications of nanoelectronics and optoelectronics based on anisotropic layered semiconductors.**

## INTRODUCTION

Two-dimensional (2D) anisotropic layered materials, such as black phosphorous (BP)[1,2], SnSe[3,4] and $ReS_2$[5,6],have attracted considerable attentions owing to their in-plane anisotropic physical properties which endow them with versatile applications [7,8], including artificial biological synapses [9], polarization-sensitive photodetectors [10] and thermoelectric devices[11]. Recently, Raman intensity [12-14], electronic transport [15,16] and optoelectronic properties [13,17,18] in monoclinic GaTe, a p-type semiconductor with a direct band gap and strong in-plane anisotropy, have been extensively studied and shown great potentials for application as high-performance anisotropic devices. Transport study demonstrates that a GaTe floating gate memory could reach a high on/off ratio of $10^7$ and a long retention time of $10^5$ s by taking the advantage of the gate-tunable strong resistance anisotropy of GaTe [15].

Generally speaking, the in-plane anisotropy of electronic structure originates from the low crystallographic symmetry. As a result, the anisotropy of energy band dispersion will show a thickness-independent character[19]. For example, the carrier effective masses of BP along armchair direction are smaller than that along zigzag direction for the thickness ranging from bulk to 2D limit[20]. In theoretical investigation, the band structure of infinitely extended bulk model is sufficient to interpret the anisotropy of band structure of BP obtained by angle-resolved photoemission spectroscopy (ARPES)[2], and the anisotropic carrier mobilities of six-layer $ReS_2$ could be understood on the basis of those of monolayer system calculated upon the deformation potential theory[5]. Previous study on GaTe shows that the highest valence band dispersion along x direction of the infinitely extended bulk model is much stronger than that obtained by ARPES, whereas a good agreement between theory and experiment is found for that along y direction[21]. Unlike BP, recent experimental study demonstrates the optical extinction ratio of GaTe is changed from ~1.3 to ~1.0 as the thickness increases[12], which indicates a quasi-isotropy of optical extinction in thicker GaTe flake. These anomalies suggest that the thickness has profound influence on the in-plane anisotropic electronic structure of GaTe. To date, there is, however, a lack of deep understanding on the physical origin of in-plane anisotropy of electronic band structure of GaTe.

In this work, we study the in-plane anisotropic electronic structure of GaTe with thickness altering from bulk to monolayer. The ARPES results show that the valence band maximum (VBM) of bulk GaTe is located at $\bar{\Gamma}$ point, and the hole effective mass along $\bar{\Gamma}$-$\bar{X}$ direction is about 3.5 times larger than that along the $\bar{\Gamma}$-$\bar{Y}$ direction. The energy band dispersion of the system is considerably weak for $k_z$ direction. The density functional theory (DFT) calculations reveal that the in-plane anisotropy of energy bands obtained by ARPES doesn't originate from the low-symmetry within the layer plane. Further analysis indicates a strong bulk-surface interaction effect plays the primary role in determining the observed in-plane band structure anisotropy. In addition, an intriguing direct-indirect-direct band gap transition is predicted from monolayer to few-layer GaTe due to strong interlayer coupling. Our findings thus demonstrate a distinct physical phenomenon, where the in-plane electronic anisotropy of a layered material is dominated by bulk-surface interaction, rather than by the crystallographic symmetry. Results of this work provide both experimental and theoretical evidences to unravel the mechanism behind the in-plane anisotropy in 2D GaTe layers, and pave the way to the design of devices making use of anisotropic layered semiconductors for future nano-electronic and optoelectronic applications.

## Results and discussion

### Geometric and electronic structure of GaTe

Bulk GaTe is a monoclinic p-type semiconductor with the C2/m space group and possesses two kinds of Ga-Ga bonds with different orientations, one of which lies almost within the layer plane and the other is oriented in the out-of-plane direction. The primitive cell of bulk GaTe is shown in Fig. 1a. The cleavage plane of bulk GaTe corresponds to the (1 1 1) plane, thus the surface Brillouin zone (BZ) is obtained by projecting the first BZ of bulk onto the (111) plane of the reciprocal space as demonstrated in Fig. 1b. After the formation of surface, the Γ and Z points in bulk BZ fold to the $\bar{\Gamma}$ point and the in-plane high-symmetry H-Z-L path of the bulk projects onto the $\bar{Y}$-$\bar{\Gamma}$-$\bar{X}$ path of the surface. The GaTe band structure revealed by ARPES using He I $\alpha$ line is shown in Fig. 1c. The VBM is located at $\bar{\Gamma}$ point, and the dispersion of highest valence band along the $\bar{\Gamma}$-$\bar{Y}$ direction is stronger than that along the $\bar{\Gamma}$-$\bar{X}$. The hole effective mass along x direction is about 3.5 times larger than that along y direction (Fig. S1a). Meanwhile, the anisotropic valence bands can also be visualized by the constant energy surfaces (Fig. 1d). They exhibit two-fold in-plane symmetry at a binding energy from 0 to 0.5 eV below VBM. The valence band spectral intensity along $\bar{\Gamma}$-$\bar{X}$ direction is elongated, which implies a stronger dispersion of the highest valence band along y direction than that along x direction. The constant energy surface reveals a four-fold in-plane symmetry as the binding energy goes down to 1.5 eV below the VBM.

**The origin of in-plane anisotropy of band structure in GaTe**

In order to evaluate whether the anisotropic energy band structure of GaTe originates from the low crystalline symmetry. The band structure of monolayer (1 ML) GaTe is calculated by using PBE functional with the inclusion of spin-orbit coupling (SOC) effect. As shown in Fig. 2a, the direct band gap is located at $\bar{X}$ point, and the highest valence band dispersion along x direction is stronger than that along y direction. These results are inconsistent with our experimental observation, indicating that the in-plane anisotropy of energy band structure of GaTe obtained by ARPES doesn't originate from the low crystalline symmetry. A higher specific surface area is expected in 1 ML GaTe as compared to its bulk counterpart and the surface effect dominates the electronic structure of the system. The discrepancy between in-plane anisotropic energy bands of 1 ML and ARPES result implies that the bulk effect plays an important role in the observed in-plane anisotropy of band structure of GaTe.

Therefore, the band structure of infinitely extended bulk GaTe is calculated. The geometric structure of bulk primitive cell is fully optimized by the optB88-vdW functional. The lattice parameters of GaTe bulk unit cell predicted by optB88-vdW functional are approximately 1.2-2.0% larger than those obtained in the experiment [14], and better than those calculated by PBE functional (Table. S1). Fig.2b shows that the band structure of bulk model possesses a direct band gap at Z point. This agrees with previous theoretical results[12,15,21]. Like 1 ML system, the hole effective mass along x direction is smaller than that along y direction for the bulk model, which doesn't agree with our experimental result (Fig. 2c). This suggests neither bulk nor surface effect could solely determine the observed in-plane anisotropy of energy bands of GaTe. Both the bulk and surface effect thus should be included in the theoretical model in order to interpret experiment.

We next perform the band structure calculation of a slab configuration with thickness of 7 ML GaTe. This configuration is built on the basis of bulk unit cell relaxed by optB88-vdW functional (Fig. S2), in which the middle five layers are fixed as bulk region while the top and bottom mono-layer are regarded as surface region, respectively. The positions of surface atoms are also optimized by the optB88-vdW functional. As shown in Fig. 2d, the PBE+SOC method predicts that the VBM of GaTe slab model is located at $\bar{\Gamma}$ point, and the highest valence band dispersion along $\bar{\Gamma}$-$\bar{Y}$ direction is stronger than that along $\bar{\Gamma}$-$\bar{X}$ direction. The hole effective mass along x direction is about 2.3 times larger than that along y (Fig. 2c. and Fig. S1d). These represent the anisotropic band structure of 7 ML slab model agrees well with our experimental result. Despite the agreement of calculation with experiment, the 7 ML slab model is not thick enough to eliminate the quantum confinement effect on the anisotropic band structure of GaTe. We then calculate the band structures for GaTe with thickness of 8 ML, 9 ML, 13 ML, 16 ML, 20 ML, 25 ML and 30 ML. The geometric structures of all multilayer systems are fully relaxed by optB88-vdW functional. The Fig. S3a shows the values of in-plane lattice constant and the band gap of the multilayer GaTe gradually converge and are close to those of bulk model as the thickness increases to 30 ML GaTe. The anisotropy of hole effective mass of 8 ML GaTe is consistent with that of 30 ML GaTe (Fig. S3b and Fig. S4), which indicates the quantum confinement effect doesn’t play a primarily role in the in-plane anisotropy of band structure of GaTe.

Moreover, Fig. S5 shows that the ARPES intensities around Fermi level ($E_F$) as a function of the momentum $k_y$ as well as the probing photon energy are rather slender (i.e. along the $k_z$ direction). From these spectra, photon energies 20 eV and 30 eV seem to be corresponding to high-symmetry points, while an inner potential $V_0$ of 16.4 eV has been obtained which qualitatively agrees with the earlier experiments[21]. In contrast to the bulk model calculations where the band structures at Γ and Z points have obvious differences, the VBM at the aforementioned two different high-symmetry points are quite close and their dispersions are similar as shown in Fig. S6. These results suggest that the $k_z$ dispersion is considerably weak and GaTe doesn't possess obvious surface states, which also agrees with early reported results[21].

Furthermore, our 7 ML slab calculation shown in Fig.2e demonstrates that the highest valence band is mainly occupied by the states from bulk region. Here, we also performed the polarization-dependent ARPES measurements. The scheme of our polarization-dependent measurements is shown in Fig. 2g. According to the selection rule, the s(LV) polarization is sensitive to probe the $p_y$ orbitals, while for the p(LH) polarization, the transition matrix element is nonzero only for photoemission originating from $p_x$ and $p_z$ orbital. Our theoretical calculations based on the 7ML slab model is shown in Fig. 2f, the highest valence band of GaTe is dominated by the $p_z$ orbital. Here, as we can see in Fig. 2h, for the highest valence band, the result of p(LH) polarization yields stronger spectral weight than that of s(LV) polarization, moreover, the valence bands between -1.5 and -2.0 eV below $E_F$ that also dominated by the $p_z$ orbital is significantly more distinguishable when probing with p(LH)-polarization. As for valence bands surrounding the $\bar{\Gamma}$ point while locates at -1.0 eV and higher binding energies is dominated by $p_x$ and $p_y$ orbitals, these band features are both pronounced under different polarizations. The distinguishing spectral weight distributions between the two measurements illustrate that the highest valence band hosts a predominantly different orbital makeup compare to band structures reside at binding energies higher than -1.0 eV.

In the discussion above, we find that the experimentally measured in-plane anisotropy of energy bands in GaTe doesn't originate from the low-crystalline symmetry and the quantum confinement effect, while also could not be reproduced by the system solely involving the surface or bulk effect. Only by considering both the bulk and surface effect does the calculated anisotropic band structure of GaTe agree with the experimental observations. Hence, there exists a strong bulk-surface interaction which dominates the in-plane electronic structure anisotropy in GaTe. The influence of bulk on the surface is attributed to the interlayer coupling. To verify this, we perform calculation within the 7 ML GaTe slab model with weakened interlayer coupling where the interlayer distances of the bulk region are 0.5Å larger than the equilibrated ones. As shown in Fig. S7a, the band structure of the system exhibits an indirect band gap. The VBM of the system is located at the point closed to $\bar{X}$ and contributed by both the $p_z$ and $p_{xy}$ orbitals (Fig. S7b), instead of $p_z$ orbitals solely. This feature is approximate to that of 1 ML GaTe in which the VBM is dominated by the $p_{xy}$ orbitals (Fig. S7c). These results indicates that the interlayer coupling will reduce the surface effect in the multilayer GaTe, which is naturally included in the bulk system. However, the significant influence of the surface effect on the in-plane band structure anisotropy in GaTe remains even though the number of layers reaches 30 ML. As shown in Fig. S3b, with the number of layers increasing to 30 ML, the hole effective mass along y direction gradually converges to that of bulk model, and a large difference between 30 ML and bulk GaTe could be found for x direction. As a result, 30 ML GaTe has an opposite in-plane anisotropy compared with that of the bulk model. This means that once the periodicity perpendicular to the cleavage plane of bulk GaTe being broken down, the highest valence band dispersion along x direction becomes much weaker than that along y direction. Therefore, the bulk-surface interaction originates from the interlayer coupling and out-of-plane periodicity breaking in GaTe.

**Band structure evolution from monolayer to few-layer GaTe**

The difference between the positions of VBM in 1 ML and 7 ML GaTe suggests a band structure evolution from monolayer to few-layer GaTe. To verify this, we provide theoretical calculations for the electronic structures of GaTe in dependence on the number of layers from 1 to 9. The geometric structures of all the systems are fully optimized using optB88-vdW functional. Table. S2 summarizes the structural parameters of few-layer GaTe. The in-plane lattice parameter of the system is non-monotonically increasing with thickness. The values of interlayer distances are in the range from 2.01 to 2.16 Å for the systems, which are much smaller than those in BP [20]. The charge density difference between 6 ML and 1 ML GaTe shows electron accumulation in the interlayer region of GaTe, indicating strong interlayer coupling in GaTe (Fig. S8). As shown in Fig. 3(a-i), the band structures calculated by PBE+SOC method for few-layer GaTe suggest the system experiences a direct-indirect-direct band gap transition.

We further analyze the orbital-projected band structures of GaTe for the number of layers from 1 to 6. As shown in Fig.S9a, both the CBM and VBM of 1 ML GaTe situated at $\bar{\Gamma}$ are derived from the delocalized out-of-plane $p_z$ orbitals. When two monolayers are stacked to form 2 ML GaTe, the VBM of $\bar{\Gamma}$ point is lifted in energy relative to that of $\bar{X}$, and the positions of CBM and VBM shift from $\bar{X}$ to the points close to $\bar{\Gamma}$ and $\bar{X}$, respectively, as a result of strong interlayer coupling. This induces a direct-to-indirect band gap transition. The 2 ML, 3 ML and 4 ML GaTe are indirect band gap semiconductors, in which the VBM and CBM are composed of $p_{xy}$ and $p_z$ orbitals (Fig.S9b-d). When the number of layers reaches 5, the VBM and CBM shift to $\bar{\Gamma}$ point and are dominated by the $p_z$ orbitals (Fig. S9e), leading to an indirect-to-direct band gap transition. As shown in Fig. 3j as well as in Fig. S1b, Fig. S10 and Fig. S11, the in-plane anisotropy of hole effective masses is reversed with the number of layers, and the ratio between hole effective masses along x and y directions reach minimum and maximum value in 3 ML and 5 ML GaTe, respectively. In contrast to the layer-dependent anisotropy of hole effective mass, the anisotropy of electron effective masses shows layer-independent feature (Fig. S1b, Fig. S10 and Fig. S11). Moreover, we conduct angle-resolved polarized Raman measurement for GaTe thin films with thickness of 3.53, 4.87 and 6.47 nm encapsulated by hexagonal boron nitride (h-BN). The optical microscopy images, Raman maps and atomic force microscope (AFM) height profiles of GaTe thin films are shown in Fig. S12. The polarized Raman spectra exhibit strong in-plane anisotropy of the Raman intensity which is consistent with that of bulk GaTe (Fig. S13 and Fig. S14). This indicates that the GaTe thin flake may be very sensitive to ambient conditions, though a monoclinic to hexagonal phase transition with the decreasing of the GaTe layer thickness has been reported before[14,22]. Our results demonstrate that the GaTe monoclinic phase is robust for the flake thickness down to 3.5 nm, and the anisotropic band structure evolution with thickness for GaTe is potentially feasible.

Previous studies show the quantum confinement effect plays an important role in the band gap evolution of layered materials[23,24]. We thus use the empirical equation $E=A/N^{\alpha}+E_{bulk}$, where E is the band gap of multilayer GaTe, N is the number of layers and $E_{bulk}$ is the band gap of bulk GaTe, to evaluate the influence of quantum confinement effect on evolution of band gap for GaTe. The fitted parameters A, α and $E_{bulk}$ are 0.70 eV, 0.79 and 0.68 eV, respectively. As shown in Fig. 3k, the band gap of GaTe decreases as the thickness increases, following the $1/N^{0.79}$ power law. The fitting exponent is smaller than the values of the infinite potential well model and BP[23], which suggests both quantum confinement and strong interlayer coupling control the band gap evolution of GaTe.

**Conclusion**

In conclusion, the in-plane anisotropic energy band structures of GaTe are systematically investigated both in experiment and theory. A two-fold symmetry valence band dispersion near VBM is found, where the hole effective mass along y direction is smaller than that along x direction. The observed in-plane anisotropy of energy bands in GaTe are the result of strong bulk-surface interaction rather than the reduced crystallographic symmetry, surface effect and quantum confinement effect. The energy band dispersion along $k_z$ direction is considerably weak in the system. As the thickness increases from monolayer to few-layer, the band structure of the system experiences a direct-indirect-direct band gap transition due to strong interlayer coupling and in turn the anisotropy of hole effective mass will be reversed. Our findings provide guidance to understanding the underlying mechanism of anisotropic electronic transport in GaTe and further improving the performance of electronic and optoelectronic devices based on the layered anisotropic semiconductors.

## METHODS

### Preparation of GaTe bulk crystals

Single crystals of GaTe used in this work were prepared via the self-flux method. We first mixed raw material with the stoichiometric ratio of Ga (purity 99.999%): Te (purity 99.9%) of 48.67:51.33 (%wt/wt), and kept the mixture at 880 $^{o}$C for about 5 hours. The mixture was then cooled down to room temperature at a rate of 1.5 $^{o}$C per hour.

### Angle-resolved photoemission spectroscopy

ARPES measurements were performed at BL13U beamline of National Synchrotron Radiation Laboratory (NSRL), and Vacuum Interconnected Nanotech Workstation (NANO-X), both using ScientaOmicron DA30L electron spectrometers. All samples were cleaved in situ and measured at 80-120 K under vacuum conditions not more than $1 \times 10^{-10}$ mbar. The angular resolution of the spectrometers was 0.3 degrees, and the combined instrumental energy resolution was better than 25 meV.

### Angle-resolved polarized Raman measurements

The h-BN-encapsulated GaTe thin films were fabricated in an mBraununiverse glove box. The thickness and morphology of GaTe thin film were obtained by a Bruker Dimension Icon AFM. Raman measurements were performed by a commercial confocal Raman spectrometer (Witec Alpha 300R).

### First-principles calculations

The density functional theory calculations were performed within the projector augmented-wave (PAW) method [25,26] implemented by using Vienna ab initio simulation package (VASP) [27,28], with a kinetic energy cutoff of 500 eV for plane-wave basis set. The generalized gradient approximation (GGA) in the Perdew-Burke-Ernzerhof (PBE) implementation [29] was chosen as the exchange correlation functional, respectively. To capture the interlayer spacing of GaTe, the optB88 [30,31] van der Waals density functional was used in the calculation. The spin-orbit coupling (SOC) effect is known to be proportional to $Z^4$ (Z is a nucleus charge) and plays an important role in the heavy element layered system [32]. Therefore, the SOC effect has been included in the band structure calculation. The Brillouin zone was sampled with Γ-centered Monkhorst-pack grid[33] of 8 × 8 × 8 and 4 × 16 × 8 for the bulk primitive and unit cell, and 8 × 8 × 1 for the surface system. A large vacuum space of at least 13 Å was used in multilayers and slab configuration of GaTe to avoid interactions between images. The atomic coordinates of bulk and surface were optimized until the maximum force of all atoms was less than 0.01 eV/Å.

**DATA AVAILABILITY**

The data that support the findings of this study are available from the corresponding author on a reasonable request.

## ACKNOWLEDGEMENTS

We thank Z. Sun for the support on ARPES measurements. This work was supported by the Science Challenge Project under Grant No. TZ2016001, the National Natural Science Foundation of China (NSFC) under Grant Nos. 11774429, 11804386, 12004429, 11974357 and U1932151, the NSAF under Grant No. U1830206 and the National Key R&D Program of China under Grant No. 2017YFA0403200, 2018YFA0306900 and 2019YFA0307800.

## AUTHOR CONTRIBUTIONS

J. D. conceived and supervised the whole work. H. W., Z. H. and M. Z. took charge of sample preparation. S. J., H. Z., B. Y., E. Z., M. Y., F. L. and S. C. carried out the high-resolution ARPES measurements. H. W. and M. Z. carried out polarized Raman measurements. K. L., X. D. and J. D. took charge of the DFT calculations. K. L., S. J., Z.

H., M. Z. and J. D. wrote the manuscript. All the authors were invited to comment on the manuscript.

K. L., S. J. and H. Z. contributed equally to this work.

**COMPETING INTERESTS**

The authors declare no competing interests.

**ADDITIONAL INFORMATION**

Correspondence and requests for materials should be addressed to Z.H. (email: vitto.han@gmail.com) or to M.Z. (email: zhumengjian11@nudt.edu.cn) or to J.D. (email: jydai@nudt.edu.cn).

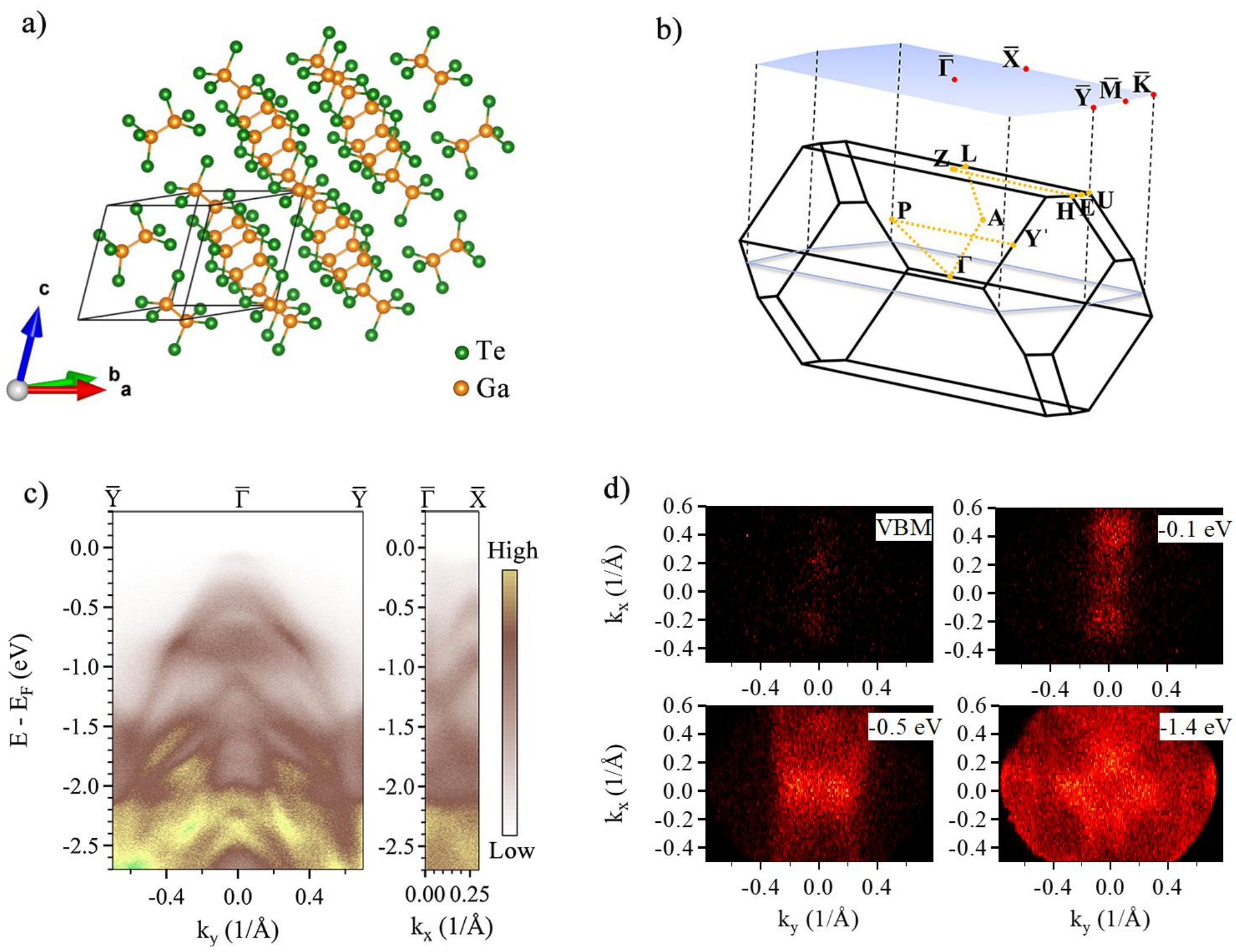

**Fig. 1. Geometric and electronic structure of GaTe. (a)** Primitive cell of bulk GaTe marked with lattice main axis (a, b, c). **(b)** First BZ of bulk and surface GaTe. The high-symmetry points of bulk and surface BZ are marked by orange and red dots,

respectively. The high symmetry path in the bulk BZ is plotted by orange dash line. **(c)** Valence bands near Fermi level along $\bar{\Gamma}$-$\bar{Y}$ and $\bar{\Gamma}$-$\bar{X}$ directions for GaTe obtained by ARPES using He I $\alpha$ line. **(d)** Constant energy surfaces corresponding to (c) with the binding energy from VBM to 1.4 eV below VBM.

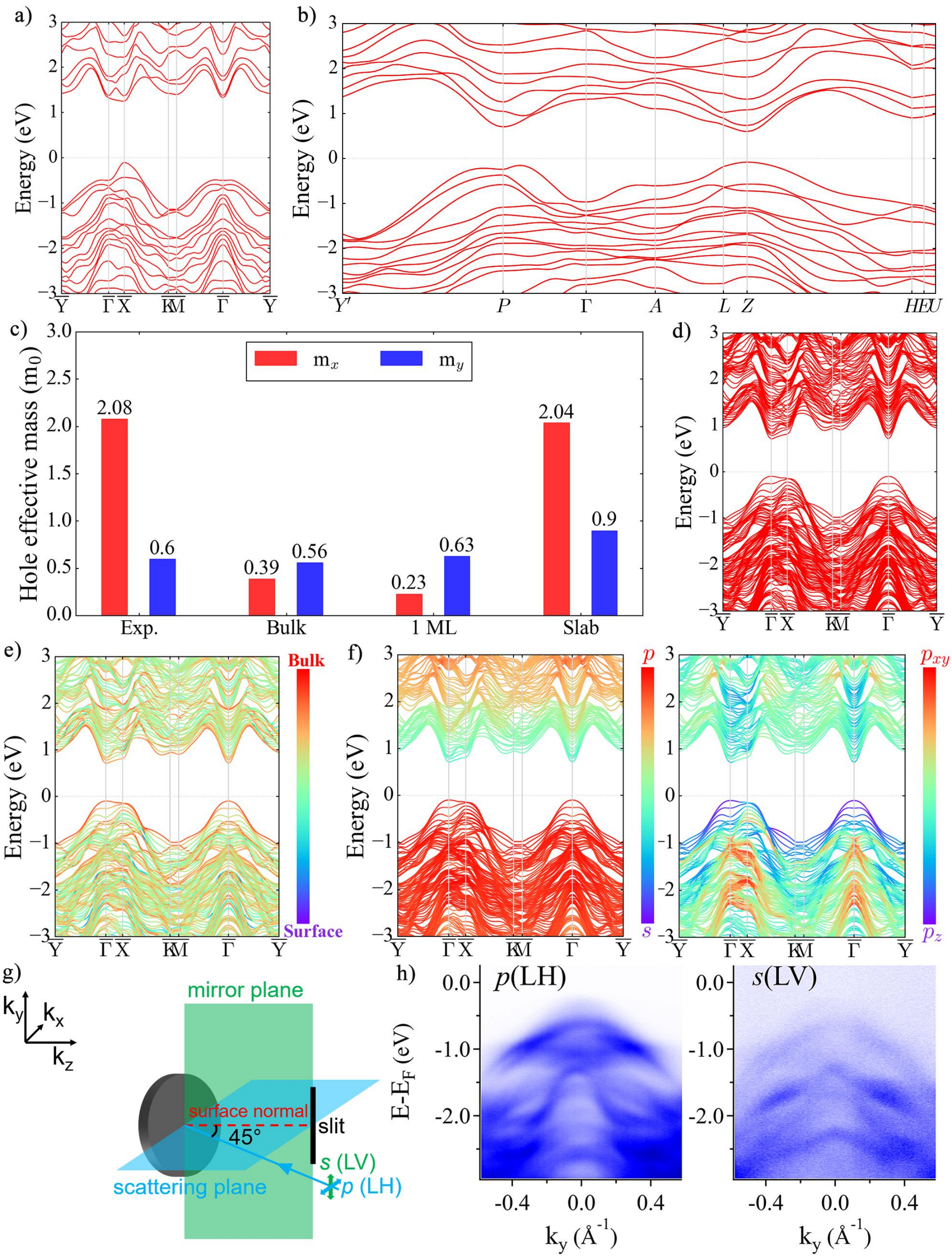

**Fig. 2. Comparison between anisotropic band structures for experiment and**

**theoretical models. (a-b)** Band structures of 1 ML and bulk GaTe calculated by the PBE+SOC method. **(c)** Hole effective masses for x and y direction for bulk, 1 ML and 7 ML slab configuration calculated by PBE+SOC method compared with experimental results in units of free electron mass $m_0$. **(d)** Band structure of 7 ML GaTe slab configuration. **(e)** Spatial-resolved band structure of 7 ML GaTe slab configuration. **(f)** Orbital-projected band structure of 7 ML GaTe slab configuration. **(g)** Scheme of polarization-dependent ARPES measurement. **(h)** ARPES spectra of GaTe obtained by using p(LH)- and s(LV)- polarized light of 21.2 eV.

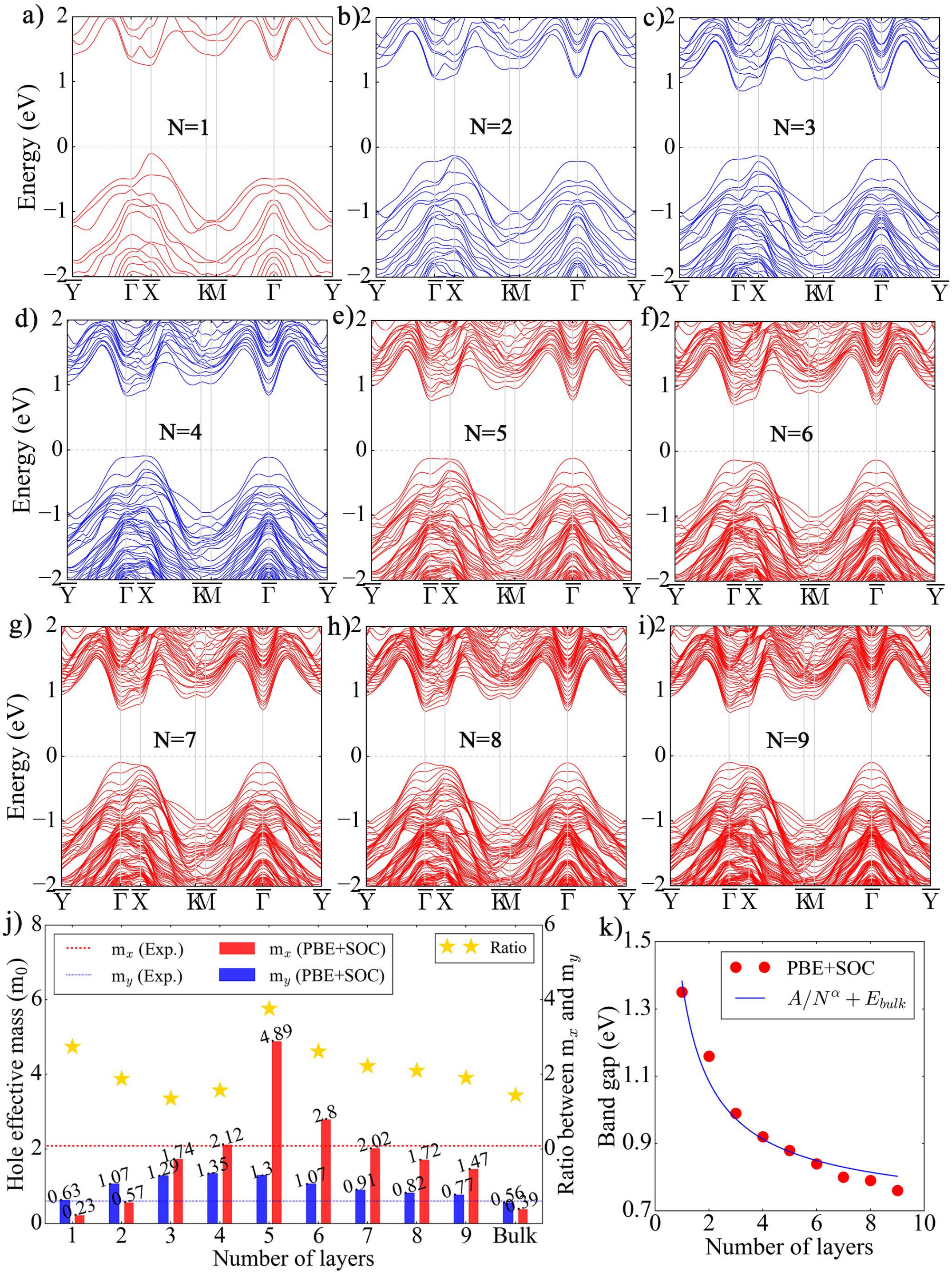

**Fig. 3. Layer-dependent band structures in GaTe. (a-i)** Band structures of GaTe with the number of layers (N) increasing from 1 to 9. **(j)** Evolution of hole effective masses along the x and y directions ($m_x$ and $m_y$) in units of free electron mass $m_0$, and the ratio between $m_x$ and $m_y$ with number of layers. **(k)** Band gap evolution as function of number of layers for GaTe. These results are calculated by the PBE+SOC method.